\documentclass[11pt]{article}

\usepackage[style=apa,natbib=true]{biblatex}
\title{Real-time Learning and Evolution in Robotic Art Installations}
\author{Sofian Audry$^{1}$, Stephen Kelly$^{2}$ \\
\small $^1$School of Media, Faculty of Communication, Université du Québec à Montréal \\
\small $^2$Department of Computing and Software, Faculty of Engineering, McMaster University \\
\small Corresponding: \texttt{spkelly@mcmaster.ca}}
\date{}

\usepackage{amsmath,amsfonts}
\usepackage{algorithmic}
\usepackage{algorithm}
\usepackage{array}
\usepackage[caption=false,font=normalsize,labelfont=sf,textfont=sf, labelformat=simple]{subfig}
\usepackage{textcomp}
\usepackage{stfloats}
\usepackage{hyperref}
\usepackage{graphicx}
\usepackage{caption}
\begin{document} 

\maketitle

\begin{abstract}
We present three robotic art installations which explore the aesthetics of adaptive behavior. Through embodied machine leaning and digital evolution, these works draw viewers into an artificial ecosystem in which open-ended novelty, trial-and-error learning, competition, and cooperation emerge in real time. Research-creation practices are examined in relation to these works, focusing on how they redefine the role of artists within a human–machine collective while examining points of convergence and divergence between artistic and engineering approaches to adaptive robotics. The systems in question use learning and evolutionary processes not as a means to optimize a specific solution, but as an aesthetic experience on its own, suggesting new modes of interdisciplinary art-science research. Finally, we discuss strategies and practices to elevate the aesthetic experience for audiences, including contexts of presentation as well as temporal and material considerations for artworks based on embodied adaptive systems.
\end{abstract}


\section{Introduction}

Robotic artworks leverage public curiosity, interaction, and surprise to provoke situated experiences of human-machine coexistence. Imagine a community of water-dwelling robots that chirp, flash, and twirl around a public pond in front of passersby, their behavior evolving as their environment changes and dusk sets on the esplanade. In an art gallery, robotic probes wriggle and hum as they crawl along fluorescent lights, slowly adapting through a coevolutionary interaction spanning a month-long exhibition. Meanwhile in a museum, visitors enter a dark room and witness an uncanny performance featuring three luminescent robots wearing silicon skins who learn to achieve simple goals in real time, hesitantly and clumsily moving through trials and errors.

These examples represent emerging forms of robotic art installations where machine learning and digital evolution build adaptive behaviors that emerge in real time. Following a decades-long tradition of robotic and artificial life art \citep{Whitelaw2004-Metacreation}, these works explore adaptive forms of "aesthetics of behavior" \citep{Penny1987-Simulation}, a paradigm where the dynamic autonomous behavior of an artwork is the primary medium of aesthetic experience. The novelty these works bring is that, rather than relying purely on classical programming and symbolic AI, they involve teleological autonomous \textit{agents} who attempt to reach a goal, often as part of a complex self-organizing system. These works align with recent artistic explorations of adaptive robotics such as Petra Gemeinboeck and Rob Saunders' curiosity-driven robots in \emph{Accomplice} (2014), or Ruairi Glynn's evolutionary dancing robots \emph{Performative Ecologies} (2020), which similarly investigate learning processes to generate aesthetic experiences through embodied behaviors.

Yet, proficient goal-seeking behavior is not the artistic objective here: rather, behavior emerges as a byproduct of the search and optimization process. These robots are not only learning to solve problems, they are \textit{learning loudly for all to experience}. 
Our motivation is to unleash these systems of embodied, adaptive behavior within immersive, experiential environment in order to cultivate critical dialogue about human-machine relationships. In doing so, we invite audiences to navigate the ambiguous territory where technological systems and living organisms meet.

Rather than superficially mimicking real-life scenarios, these works introduce key attributes of biological living systems such as modularity \citep{callebaut_2005}, lifetime learning and adaptability \citep{watson_2024}, and self-replicating information \citep{adami_2024}. Thus, instead of designing machines that act like animals and accurately reproduce animal-like sounds, the objective here is to develop complex mechatronic agents that are made from multiple building blocks (modularity), capable of adapting their behavior, physical structure, and sonic properties based on interaction with their environment (adaptability), and in some cases reproduce by making digital copies of themselves with variation (self-replicating information). In the case of evolutionary processes, our design goal for these artworks is not to simulate evolution, but rather to create \textit{instantiations} of evolution \citep{pennock_2007}. The behavior of these works undergoes actual Darwinian evolution: iterative variation, selection, and inheritance operating on a population of digital organisms. This follows a methodology in digital artificial life that seeks to discover new knowledge about the natural world by studying such nonbiological creations.

We approach these projects through a \emph{research-creation} framework, a research paradigm where creative practice functions as a mode of inquiry, generating experiential and embodied knowledge ~\citep{Chapman2012-ResearchCreation,Loveless2019-How,Paquin2020-Degager}. While traditional scientific research produces knowledge through theoretical abstraction and empirical validation, research-creation generates knowledge through situated material encounters and sensory experience, therefore producing experiential, embodied, and sensible forms of knowing. A central contribution is therefore the mere fact that these installations \textit{exist}, and their existence has the potential to reshape how people perceive and relate to adaptive artificial agents.

In this paper, we present three robotic art installations that incorporate adaptive behavior. We then discuss practice-based considerations of creating such works, looking at how they impact the role of the artists and considering the differences and similarities with engineering approaches. Finally, we expose some considerations related to the aesthetic experience from an audience perspective, including contexts of presentation as well as temporal and material design.

\section{Robotic Art Installations}

In order to ground this discussion in concrete use cases, we describe three art installations employing diverse types of autonomous robotic agents adapting in real-time through interaction with their environment. These works use customized algorithms based on swarm optimization, genetic programming, and reinforcement learning.
While these algorithms are commonly associated with optimization, we are primarily interested in how the agents search a space of possibilities and how they build new behavior through interactive experience, rather than how well they optimize any single solution.

These works represent radical refusals of the race for more data and more computing resources commonly seen in the 'big science' acceleration of AI and machine learning. Instead, they leverage: 
\begin{itemize}
\item Low cost, low power embedded computers and simple \textit{shallow learning} representations that don't rely on specialized hardware acceleration (e.g. GPUs);
\item Small, local, and ephemeral data derived from egocentric sensors as opposed to \textit{big data} sets;
\item Embodied algorithms that unfold through the movement of human-scale entities at human-perceptible timescales, as opposed to highly parallelized code running in data centres.
\end{itemize}
In their own way, these works all explore what we can learn by building and sharing robotic systems detached from the goals of modern AI economies, focusing instead on embodied machine learning and evolution for cultural exploration.

\subsection{Vessels (2015)}

\begin{figure}[!htbp]
    \centering
    \includegraphics[width=.95\linewidth]{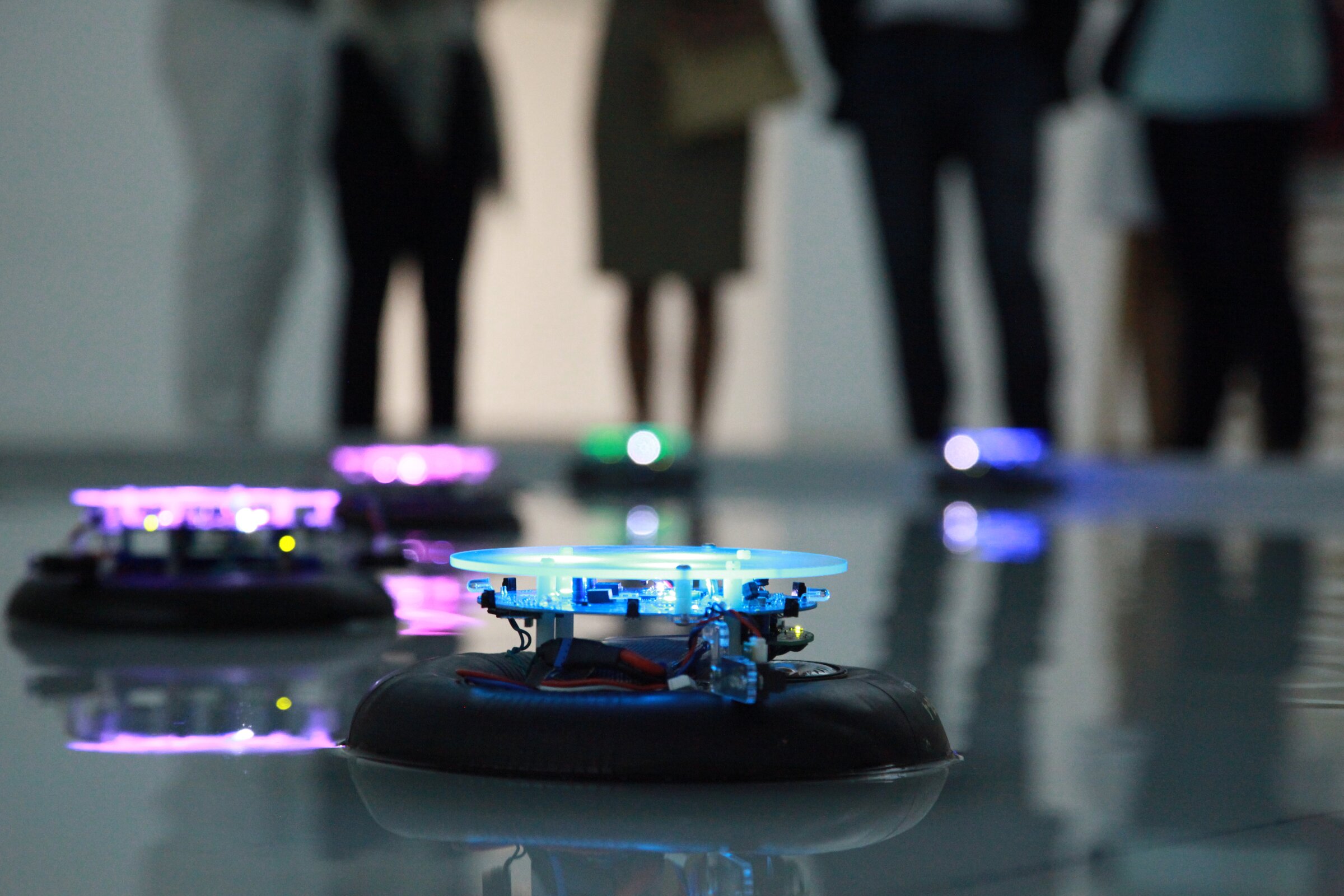}
    \caption{\textit{Vessels} by Sofian Audry, Stephen Kelly and Samuel St-Aubin (2015). Photo credit: Beatriz Orviz. Video link: \url{https://vimeo.com/758502989}}
    \label{fig:vessels-main}
\end{figure}

\textit{Vessels} consists of a small swarm of between 10 and 20 autonomous water robots who interact with each other and their environment, Figure \ref{fig:vessels-main} (Video link: \url{https://vimeo.com/758502989}). Their collective behavior resembles the social interactions in a community of living creatures. The work prompts audiences to interpret the robots' behaviors, attribute intentions to their actions, and perceive how the group dynamic reveals otherwise imperceptible qualities of the environment they share with the robots. \footnote{The artists and exhibition docents have observed visitors and passersby spontaneously generate narratives about the robots' social dynamics, ascribing relationships such as friendship, courtship, and enmity to their behaviors. People are often compelled to touch, push, or even kick them. Children demonstrate particular facility for this imaginative projection, often constructing complex stories around the autonomous agents.}

While moving on water, each robot collects and interprets data from different environmental conditions such as water and air quality, temperature, ambient light, and sound. Since each agent has only one of such sensors, they depend on communicating with each other to form a complete awareness of their environment. However, the robots do not directly exchange sensor data: instead, they communicate through behaviors and interactions. For example, an increase in temperature sensed by one agent may cause it to act more aggressively, with erratic or irrational (random) movements. This change in behavior will influence its neighbouring agents, who may respond with changes to their own behavior, and so on.

Over time, a distributed collective behavior specific to the environmental characteristics of the presentation site emerges from the agents’ interactions. The work thus acts as responsive laboratory that dynamically represents hidden features of the urban ecosystem by displaying emergent social behaviors, offering the viewers a new perspective on their living milieu.

The artistic intention behind the piece is to develop a form of adaptive, distributed choreography: a constantly evolving dance unfolding without a central conductor. This dynamic interplay fosters a sense of aliveness, inviting the audience to more intimately relate to the work, identify with it, and ultimately reflect and question their own relationship with their living environment. 

Technically, these goals were approached by combining two software components: 1) A formal, rule-based, goal-oriented system known as a Behavior Tree that runs the low-level decisions such as avoiding obstacles or fostering socialization; and 2) a custom genetic algorithm that generates personalized manifestations of behaviours.

The Behavior Tree (BT) is a tree-like, agent-based structure that provides a high-level, stable interface that manages priorities and sequences of actions. Originally developed in the field of video game design as a surrogate to Hierarchical Finite State Machines (HFSM) in the modeling of agent behavior~\citep{Isla2005-Handling}, it has been applied recently to robotic control~\citep{Marzinotto2014-Unified} where it can be considered as a flexible alternative to subsumption architecture.\footnote{Subsumption architecture is an approach to robotic control developed by Rodney Brooks in the mid--1980s, in response to some of the inadequacies of 'Good Old-Fashioned AI'. Whereas traditional AI approaches to robotics entailed hand-cfrafted internal representation of the world, subsumption architectures directly couple observations to actions in a layered control system. Using a bottom-up approach, the engineer iteratively adds control layers, starting with the most low-level rules and refining them while moving into higher-level control routines.~\citep{Brooks1986-Robust}}

Agent behavior evolves over time using a custom genetic algorithm that searches for diverse yet meaningful behaviors (see Figure \ref{fig:EvoRL} for a simple schematic). Each robot’s “personality” is encoded as a binary DNA sequence, capturing parameters that govern how the agent responds to sensor readings through LED animations, sounds, and motion preferences. When agents encounter each other, they perform a \textit{recombination} step which mixes their code. A \textit{fitness function} scores each new agent based on minimizing changes to its original DNA while maximizing novelty of its new behavior. Finally, a \textit{selection} algorithm stochastically filters out agents with the lowest fitness. This subverts the standard use of genetic algorithms for optimization, instead performing an evolutionary search for “family resemblances” that dynamically balance familiarity with novel responsive adaptation to shifting environmental inputs. What emerges is a population whose coherence comes from the iterative circulation and evolution of \textit{information}, encoded by binary strings and expressed as behavioral patterns.


\subsection{Open-ended Ensemble (2016)}

\begin{figure}[!htbp]
    \centering
    \includegraphics[width=.95\linewidth]{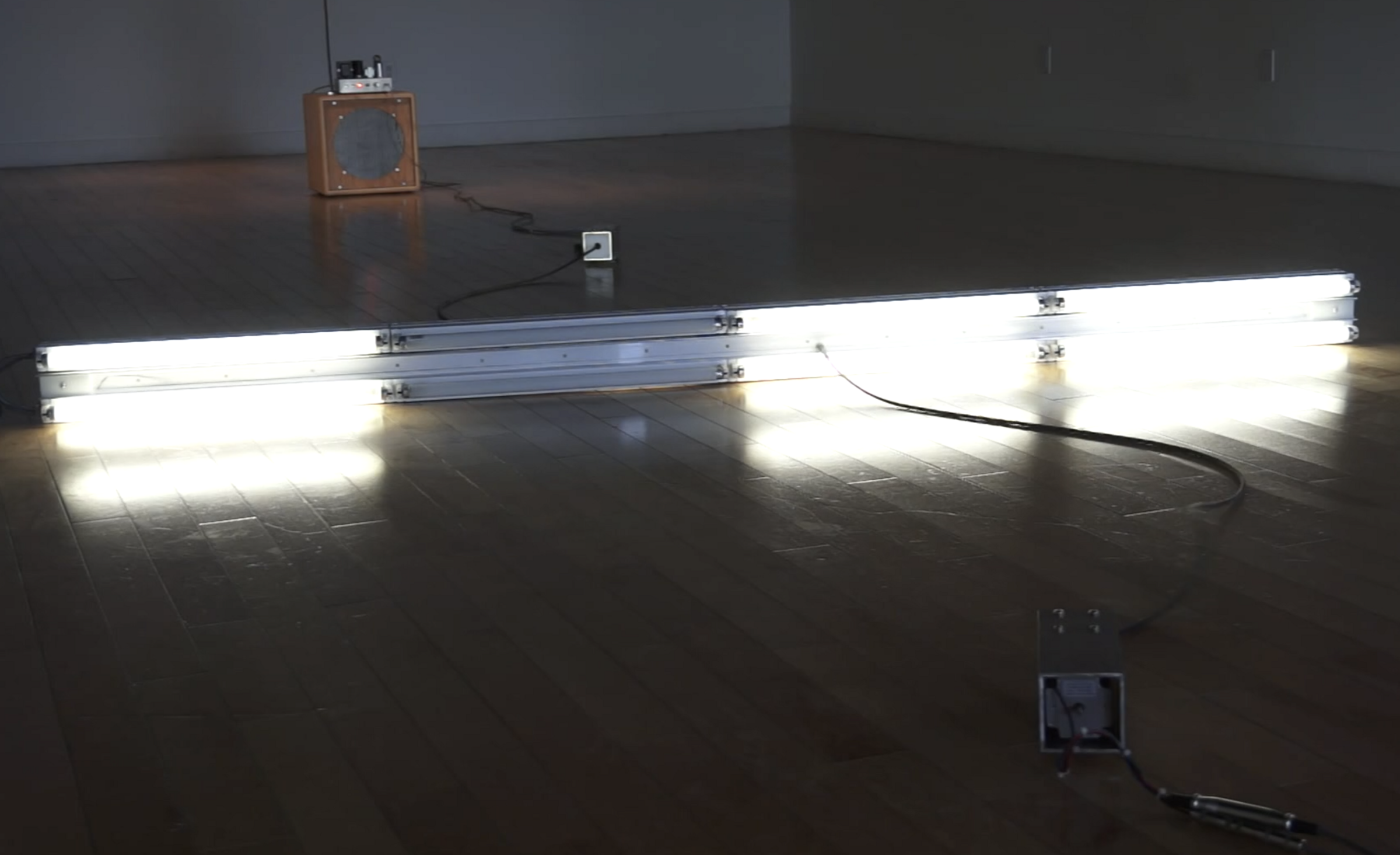}
    \caption{\textit{Open-Ended Ensemble} (Competitive Coevolution) by Stephen Kelly (2016). Photo credit: Caitlin Sutherland. Video link: \url{https://vimeo.com/194200199}}
    \label{fig:oee-wide}
\end{figure}

\textit{Open-Ended Ensemble} is a mechatronic sound installation in which agents coevolve in competitive interactions with each other, Figure \ref{fig:oee-wide} (Video link: \url{https://vimeo.com/194200199}). A pair of high-voltage sound amplifiers slowly learn to maneuver a robotic probe to the weakest region of electro-magnetic radiation along a fluorescent light fixture. Strong radiation from lights induces noisy signals in the probes and causes the amplifiers to hum, layering sympathetic tones over the familiar drone of fluorescent lighting.

An evolving population of computer programs are in control of each probe's adaptive behavior. Their intrinsic goal is to coax its movement away from the source of radiation and into silence. Meanwhile, a separate population of programs controls which bulbs are active on the light fixture. This population prefers the drone, and their selection favours behaviors which learn to trap the probe in regions of strong radiation. In this way, the robotic probes and light fixtures are interacting agents whose fitness is coupled by opposing objectives. An arms race ensues as the two competing forces interact and coevolve, akin to predator/prey relationships in nature.

In the genetic algorithm described for evolving agent behaviors in \textit{Vessels}, individuals in the evolving population are represented as binary strings. This genetic framework is extended in \textit{Open-Ended Ensemble} using \textit{genetic programming}, a special kind of genetic algorithm in which the individuals being evolved are represented as computer programs. These programs are active, information-processing devices which interact with the outside world by consuming inputs (electro-magnetic radiation) and producing outputs (control signals for the probes).

Agents in the \textit{Open-Ended Ensemble} are interacting with simple, conflicting goals in a never-ending game. Their environmental sensor readings are inherently noisy, obtained entirely from the coupling of a bare-bones magnetic probe to a fluorescent light fixture. Their control of the probe is imprecise and clumsy. As a result, the agents are navigating their world with partial information and limited motor control. A volatile model ecosystem emerges in which each device achieves intermittent success and failure as their interaction unfolds over days or weeks.

The goal of \textit{Open-Ended Ensemble} is to provide an experiential environment for audiences to contemplate tensions at various timescales of life: 1) The central tension in evolution produced by the continual interplay between diversity-producing operators such as mutation and recombination, with selection operators that drive the search toward well-adapted organisms (see \textit{Evolution} loop in Figure \ref{fig:EvoRL}); 2) The social tension between species forced to interact with competing objectives; and 3) The subjective tension experienced by each agent, stemming from cyclical frustration and satisfaction implicit in trial-and-error learning throughout one's lifetime (see \textit{Learning} loop in Figure \ref{fig:EvoRL}).

The nature of coevolutionary interaction between agents has significant impact on the character of these works. The competitive scenario produces a confrontational and menacing battle that colors the viewer experience. One visitor remarked how they "love the installation, but also hate it", referring to simultaneous feelings of empathy for simulated living agents, and emotional stress from witnessing the futility of an endless zero-sum game enfolding in the gallery. 



\subsection{Morphosis (2025)}

\begin{figure}[!htbp]
    \centering
    \includegraphics[width=.95\linewidth]{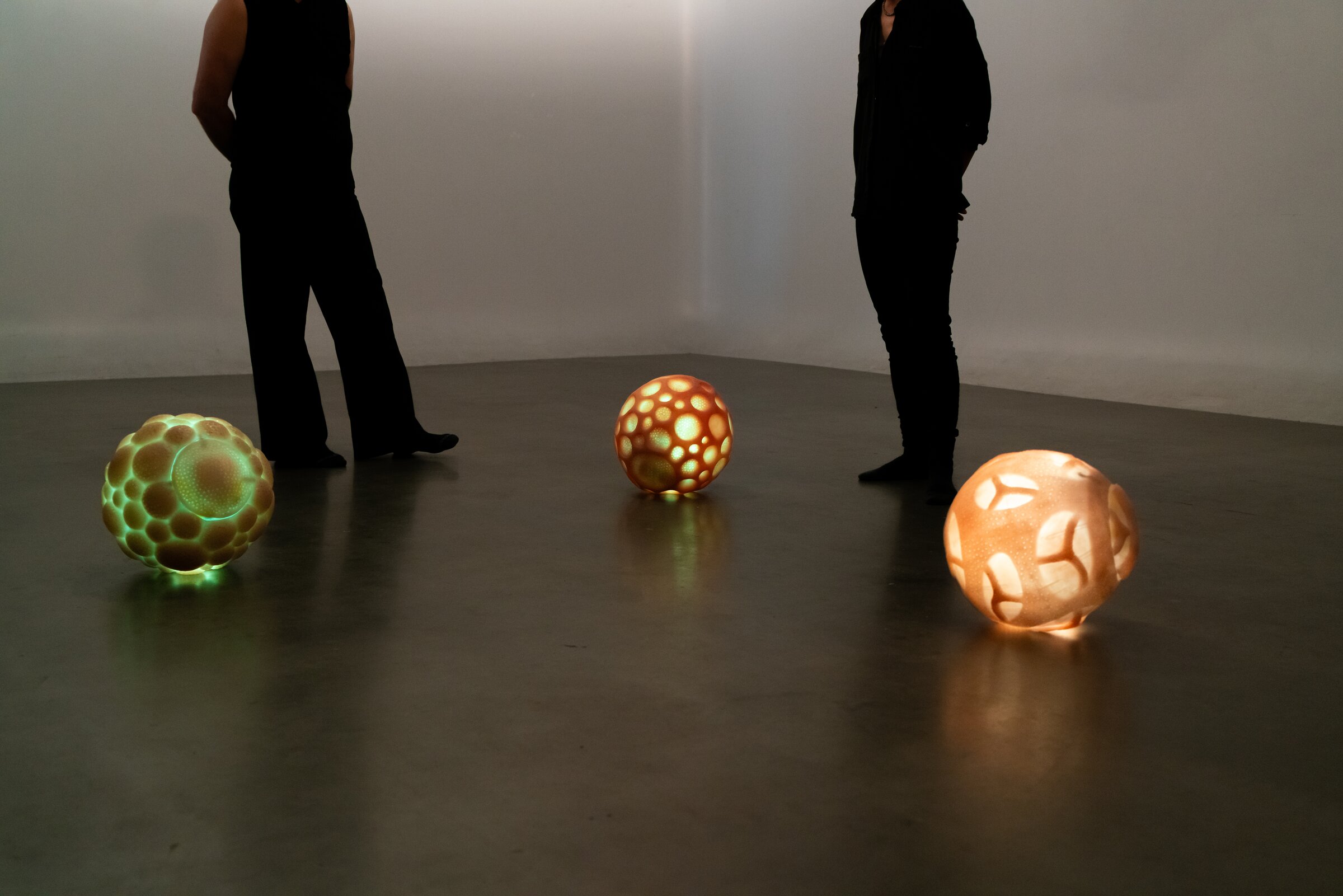}
    \caption{\textit{Morphosis} by Sofian Audry and Rosalie D. Gagné (2024). Photo credit: Léa Martin. Video link: \url{https://vimeo.com/1035428441}}
    \label{fig:morphosis-main}
\end{figure}

\textit{Morphosis} is an installation featuring three autonomous spheroid entities with distinctive morphologies, whose behavior is driven by reinforcement learning, Figure \ref{fig:morphosis-main} (Video link: \url{https://vimeo.com/1035428441}). These synthetic creatures navigate their environment, showcasing the machine learning process as an embodied and emergent phenomenon.

Visitors are invited to observe these distinctive robotic agents as they clumsily attempt to achieve simple goals such as tilting rhythmically, standing still, or moving toward specific locations. Powered by the same machine learning algorithm, each entity nevertheless encounters distinct challenges due to its specific physical form, which in turn contributes to shaping a unique behavior. As they learn before the audience, their imperfect decisions and tentative movements underscore fragility and vulnerability, offering visitors an embodied experience of machine learning that is intended to foster a sense of connection and empathy.

Guided by a reinforcement learning algorithm~\citep{Sutton1998-Reinforcement}, the entities learn and adapt their behavior through trial and error rather than being programmed to execute specific tasks. The rewards and errors are represented as greenish (positive) or reddish (negative) hue glowing through their skin. This feedback is accompanied by a real-time graphical display that visualizes their perceptions and progress, offering viewers a glimpse into the agents’ subjective experience as they adapt to their environment.

The piece is presented as a 15–20 minute performance during which the agents enact a sequence of adaptive behaviors. Each behavior is structured around a particular combination of observations (e.g., angular velocity, position, speed, distance to a target), available actions (e.g., speed and steering), and a reward function. The beginning of each behavior is marked by a short title projected onto the gallery wall, while visitors are provided with a handout listing the sequence of behaviors and their corresponding titles, allowing them to follow the progression of the performance much like a theater or concert program.

Examples of behaviors include "Being still" in which the agents are negatively rewarded for moving; "Rocking without rolling", where they are positively rewarded for generating internal motion while being penalized for horizontal displacement across the floor, typically resulting in a "rocking" motion; and "Being curious", in which the reward function is an intrinsic reward based on artificial curiosity~\citep{Schmidhuber1991-Curiosity}, encouraging the creatures to seek situations in which the outcomes of their actions are surprising.

\textit{Morphosis} challenges conventional understandings of artificial intelligence as purely efficient or functional by focusing on the process rather than the objective. It reveals the subtle and evolving nature of machine learning as a dynamic interplay between form, behavior, and environment. The work thus offers a space for reflecting on the boundaries between human and non-human, encouraging the exploration of new possibilities to co-exist with autonomous artificial systems.

The agents are driven by a lightweight Q-learning algorithm~\citep{Watkins1992-Qlearning,Sutton1998-Reinforcement} operating in real time on low-dimensional state representations. These state spaces are carefully tailored to reflect specific behavioral modalities, such as orientation or movement. Rather than relying on large-scale simulation or deep learning, the system emphasizes direct interaction with the physical world, using tabular Q-values updated through live feedback, and tiling to discretize continuous inputs. This enables each agent to learn simple tasks over a few minutes in ways that are shaped as much by their morphology and material contingencies as by the algorithm itself.
In contrast to engineering-driven approaches that prioritize efficiency or task completion, the design-driven approach focuses on unfolding the learning process itself as an embodied, situated, and visible phenomenon. 
The value of the system resides not in solving complex problems, but in producing behaviors that are often hesitant, imperfect, or even unsuccessful—revealing learning as an open-ended negotiation with the environment, and allowing viewers to witness machine adaptation in its most fragile and expressive form.

\textit{Morphosis}, \textit{Open-Ended Ensemble}, and \textit{Vessels} challenge conventional understandings of artificial intelligence as purely efficient or functional by focusing on the learning process itself rather than its outcomes. They reveal the subtle and evolving nature of machine learning as a dynamic interplay between form, behavior, and environment.

\section{Creative Process}

Using machine learning and evolutionary computation to design robotic behaviors for artistic purposes significantly transforms creative methodologies. Both conventional artistic approaches to behavior aesthetics and established engineering methods for adaptive robotics must be reimagined due to the complex interplay between artists, algorithms, machines, and installation environments. In this intricate process, the creative agency is distributed among both human and non-human participants ~\citep{Glaveanu2014-Distributed}, challenging traditional notions of authorship and control. In this section, we examine this transformation through four interrelated dimensions: first, a perspective on creative algorithms; second, how the creative autonomy of algorithms reconfigures the human-machine collective; third, how embodied machine learning resists simulation and demands situated material engagement; and fourth how interdisciplinary research-creation generates knowledge that bridges artistic and scientific practices.

\subsection{Creative Algorithms}
Broadly speaking, the main challenge in machine learning and evolutionary computation is to allow a machine to do things which it has not been explicitly programmed to do. Autonomous learning empowers robotic agents with predictive control of their situation. They perceive the environment through sensors, learn internal world models, and use these models to take actions that change their situation (see inner \textit{Learning} loop in Figure \ref{fig:EvoRL}). When the artwork is a complex adaptive system, creativity becomes a distributed collective process. How should artists position themselves when they only have partial control? 

Creativity is a process that generates something that is both new and of value \citep{boden_1991}. By this definition, evolution is a wildly creative process that is responsible for the spectacular diversity of species in the biosphere. Interdisciplinary research in \textit{open-ended evolution} suggests a formal link between evolution and the creative process by examining how evolutionary systems (computational and biological) are capable of generating unbounded novelty and complexity over time \citep{soros_2024}. Open-ended systems never settle into a single stable equilibrium. At first, this emphasis on perpetual change, novelty, and surprise seems counter to common engineering applications of evolution for direct optimization or problem solving. However, in concert with this open-ended \textit{novelty search} \citep{lehman_2011}, selection mechanisms guide the search by ascribing value to individual behaviors and determining which candidates are allowed to reproduce. The resulting search algorithms, broadly known as \textit{Quality-Diversity} \citep{pugh_2016}, are powerful optimization tools precisely because they implement novelty \textit{and} value search together in a way that satisfies common definitions of creativity. 

On a shorter timescale, individual lifetime learning also empowers robotic agents with predictive \textit{and} creative control of their situation. They perceive the environment through sensors, learn internal world models, and use these models to take actions that change their situation in both exploratory (novel) and goal-oriented (valuable) ways (see \textit{Learning} loop of Figure \ref{fig:EvoRL}). Thus, the active exploration/exploitation behaviour implicit in trial-and-error reinforcement learning also instantiates core elements of creativity.

\subsection{The Human-Machine Collective}
As evolutionary and learning algorithms exhibit increasing creative autonomy, the artist's role shifts from direct control to indirect guidance. By carefully choosing materials, hardware design, environment, algorithms, models, and hyperparameters, they "set the table" for a behavior to unfold autonomously. This requires management of several conflicting objectives: 1) Creating meaningful experiences for a human audience--which may include the artist as observer; 2) Allowing for machine agency, autonomy, and the emergence of surprise; and 3) Supporting artistic control by creating a system that is manageable for the artist to co-create with the robot(s).

The concept of \textit{Umwelt}, introduced by biologist Jakob von Uexküll, is useful to examine the artistic challenge at play \citep{vonUexkull1957-Stroll}. Rather than viewing nature as a single, shared reality, von Uexküll argued that each species lives in its own meaningful world, or \textit{Umwelt}, structured by its biological needs and capabilities. It describes the subjective perceptual world of an organism: the unique way in which it experiences and interprets its environment, shaped by its sensory and cognitive capacities.

Though originally developed for biological organisms, applying \textit{Umwelt} to the case of artificial agents ~\citep{Froese2009-Enactive,Yazici2018-Relevance} proves useful in addressing the design problem faced by artists working with adaptive robotics. These artists must create conditions suitable for robotic learning, while ensuring that their behavioral development can be perceived by human audiences. This requires careful design of both sensory and expressive features that support the experiential encounters of two worlds: the human's and the machine's. Artists thus craft specific sensory and functional affordances on each side, shaping the conditions for significance and interpretation. Within this shared space, the aim is not purely observation nor control, but the co-creation of new experiential possibilities by both artists and machines.

Artists working with machine learning agents grapple with the tension between their own intentions, creative control, and authorship on one hand, and the machine's computational autonomy and generative capacity on the other. This conundrum is a core dimension of computational creative practices such as generative and robotic art~\citep{Colton2012-Computational,McCormack2019-Autonomy}.
At one extreme, some view the machine as a tool devoid of agency or creativity, akin to a painter's brush. At the other end, some advocate that these machines are fully autonomous, leaving their human creator as the initial inventor or meta-creator. Artist Theo Jansen, for example, works towards the ultimate objective of his robots living indefinitely and unassisted in their natural (beach) environment ~\citep{herzog_2014}.
Between these extremes, artists have articulated various middle-ground postures in which machines are considered co-creators, inspirational muses, creative assistants, or even equal collaborators ~\citep{Audry2021-Art}.
Evolutionary artist William Latham describes his role as a gardener, carefully selecting and pruning a space of possibilities generated by the computer ~\citep{Kemp_1998}. Artist Rita McKeough, having a background in performance art and music, incorporates robotic agents as surrogate or collaborative performers in works that draw attention to the impact of creeping urbanization and resource extraction on plants and animals \citep{sherlock_2018}.

None of these metaphors fully captures the alien yet familiar nature of machine learning behaviors, prompting artist Memo Akten to propose the more fitting analogy of taming a wild horse ~\citep{Audry2021-Art}. This concept puts both human and machine on equal grounds, but puts the lead in the hands of the human who needs to develop an intuitive and embodied understanding of the beast, to develop a common language in order to achieve its goals. This process echoes the French concept of \textit{apprivoisement}, an iterative act of seduction and trust-building not aimed at subjugation and control, but fundamentally at building a relationship with an uncanny Other ~\citep{Audry2024-Choreomata}.

\textit{Vessels}, \textit{Open-Ended Ensemble}, and \textit{Morphosis} incorporate aspects of all these approaches. Although artists are architects of the installation experience, the works are designed to create an opportunity for the artist \textit{and} audience to engage in the search and discovery process within an artificial ecosystem, and to contemplate the volatile intersection of natural and manufactured worlds.

\subsection{Embodied Machine Learning}

Adaptive behaviors emerge at the meeting point between a machine learning algorithm and a physical body, each robot having a distinct morphology to which it must adapt. Making an agent's learning process perceptible to viewers requires precise manipulation of this behavioral matter to allow for the expression of indeterminate movements and reactions. 
However, a key limitation of using real-time machine learning in these embodied, robotic artworks lies in the near impossibility of simulating the systems involved.

\begin{figure*}[!t]
\centering
\subfloat[]{\includegraphics[width=.9\linewidth]{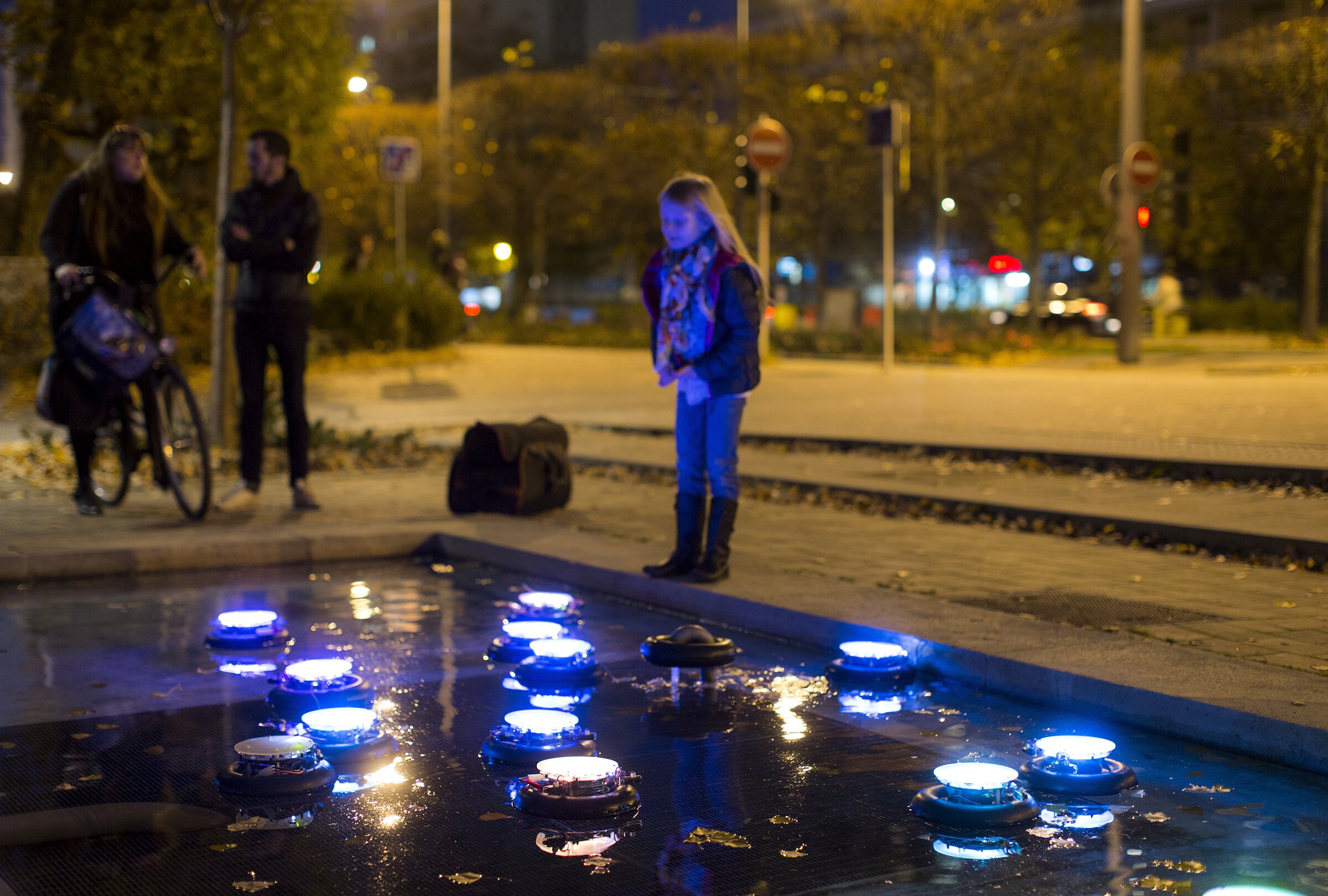}%
\label{fig_first_case}}
\\
\subfloat[]{\includegraphics[width=.9\linewidth]{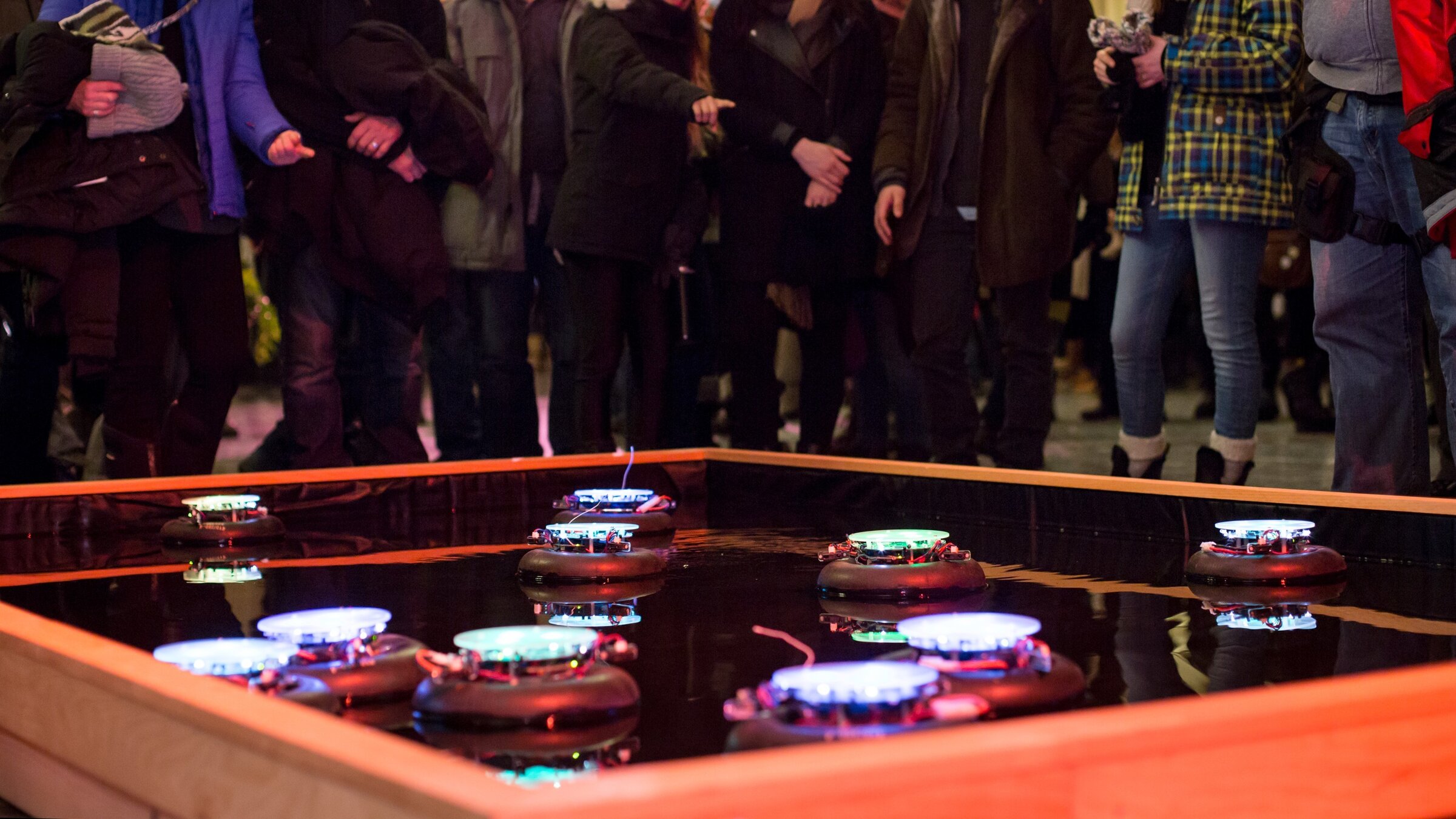}%
\label{fig_second_case}}
\caption{\textit{Vessels} by Sofian Audry, Stephen Kelly and Samuel St-Aubin (2010-2024) is shown in different contexts. (a) Outdoor presentation at L'Ososphère Festival 2015 (Strasbourg, France). Photo credit: Philippe Groslier. (b) Indoor presentation at Montréal City Hall during Nuit Blanche 2016 (Montréal, Canada). Photo credit: Catherine Aboumrad.}
\label{fig_sim}
\end{figure*}

Simulation is fundamental in engineering-driven approaches to robot training, allowing agents to be trained over thousands of iterations, safely and efficiently, before being deployed in the real world~\citep{Zhao2020-SimtoReal}. However, these simulations typically depend on tightly controlled hardware and environments, which stand in stark contrast with artistic practices. Artists seek to create unique and distinctive works, preferring iterative approaches where the robot design evolves throughout the creative process and often incorporates found materials and improvisation. Surprising and risky results are often provocative in ways that serve artistic intent. Furthermore, the environments in which these robots adapt might differ significantly for each presentation. For example, \textit{Vessels} has been shown in both gallery and outdoor contexts (see Figure \ref{fig_sim}), with different configurations of basins, quantity of robots, and environmental conditions.

Simulation-based pretraining cannot capture the embodied experience of the real system in diverse settings, nor anticipate what parameters will ultimately matter~\citep{Audry2020-Behaviour}. More critically, real-time learning is not merely a means to an end in these works, but constitutes their aesthetic and conceptual core: even failure can yield meaningful aesthetic experiences. Optimizing this process away through pretraining risks erasing the very qualities that make the system compelling.

Without simulation, the learning and adaptation process must happen within the physical environment of the piece, often through long experimentation. In-person artistic residencies become essential, especially in collective settings. For \textit{Vessels}, each residency required constructing water basin; while during COVID-19 a \textit{Morphosis} robot had to be shipped abroad so a collaborator could improve reinforcement learning algorithms in direct contact with the hardware.

Moreover, because these systems learn under changing conditions, reproducing specific behaviors is nearly impossible: tuning parameters over several sessions may yield a desired effect one day, only to shift the next. The metamorphic nature of adaptive and evolutionary artworks means behavior continually oscillates between stability and transformation ~\citep{Audry2020-Behaviour}.

These constraints invite a mode of making where meaning arises not from prediction or planning, but from sustained engagement with the material and temporal dynamics of the work itself, where adaptation may unfold over hours, days, or longer.

\subsection{Research-Creation and Artificial Life} 

The authors of this text are both transdisciplinary artist-researchers with formal education in science \textit{and} the arts. The works presented here emerge from a research-creation approach in which artistic, technical, and theoretical inquiry develop simultaneously through iterative and critically-informed processes of making \citep{Chapman2012-ResearchCreation,Loveless2019-How,Paquin2020-Degager}. Rather than treating artistic production as a secondary mode of dissemination for scientific research, research-creation positions the creative process and the artwork itself as sites of knowledge production.

Artificial Life has long maintained strong connections with artistic practice \citep{Penny2009-Art,Whitelaw2004-Metacreation,Wu2024-Survey}. Since the 1990s, many artists and researchers such as Richard Brown, France Cadet, Erwin Driessens and María Verstappen, Jon McCormack, Simon Penny, Christa Sommerer and Laurent Mignonneau, Karl Sims, Nell Tenhaaf, and Takashi Ikegami have explored emergence, artificial ecologies, evolutionary processes, and embodied interaction through installations, robotics, and generative systems \citep{Sommerer1999-Art,Tenhaaf1998-Art,Brown2001-Biotica,McCormack2009-Evolution,Penny2010-Twenty,Ikegami2013-Design}. In his foundational book \textit{Metacreation: Art and Artificial Life}, Mitchell Whitelaw framed Artificial Life art as a distinct artistic field centered on generative, autonomous, and evolutionary systems capable of producing behaviors and forms beyond direct authorial control \citep{Whitelaw2004-Metacreation}. Dedicated exhibitions, publications, and initiatives such as the VIDA Art and Artificial Life International Awards (1999--2014) further contributed to defining and disseminating Artificial Life art internationally \citep{Tenhaaf2008-Art}.

The projects presented in this paper build on this tradition through robotic installations in which learning, evolution, and embodiment are treated simultaneously as technical processes and aesthetic material. These artworks have been exhibited internationally in art institutions and festivals while also contributing directly to scientific research. The genetic programming algorithms developed for \textit{Open-Ended Ensemble} were later extended and published in scientific contexts for evolutionary reinforcement learning \citep{kelly_2018b} and robot control \citep{kelly_2023}. Similarly, machine learning systems developed for \textit{Morphosis} contributed to subsequent research on behavioral machine learning \citep{Audry2020-Behaviour}. However, the primary objective of these works is not the optimization of predefined tasks, but the creation of embodied artificial ecologies in which adaptation, open-ended interaction, and emergent behavior become observable aesthetic experiences. In this sense, the installations operate simultaneously as public artworks and as situated experimental systems through which audiences encounter adaptive robotic behaviors unfolding in real time.

Although science, engineering, and art have often been institutionally separated since the industrial revolution, Artificial Life has historically operated across these boundaries through synthetic and exploratory approaches to understanding adaptive behavior \citep{Penny2009-Art}. Recent decades have also seen widespread democratization of robotics and embedded technologies through open-source hardware, low-cost electronics, online documentation, and hacker and maker communities. These developments have enabled creative practitioners, even without formal engineering training, to design and construct increasingly sophisticated autonomous systems, blurring the line between artist, engineer, and researcher \citep{butts_2022}.

Such accessibility expands the potential for interdisciplinary collaboration while also supporting alternative approaches to engineering design beyond problem solving and constraint satisfaction, including improvisation, experimentation, play, and societal critique. Research-creation provides a strong research paradigm for such work by recognizing the construction of systems and artifacts as a legitimate mode of scholarly inquiry \citep{Chapman2012-ResearchCreation,Loveless2019-How,Paquin2020-Degager}. This orientation shares core epistemological and methodological principles with Artificial Life, where synthetic approaches seek to understand adaptive behavior through the creation of embodied and situated systems exhibiting lifelike properties. In both cases, knowledge emerges not exclusively through abstract analysis or controlled experimentation, but through iterative processes of design, interaction, observation, and material engagement with complex systems.

\begin{figure}[!htbp]
    \centering
    \includegraphics[width=.95\linewidth]{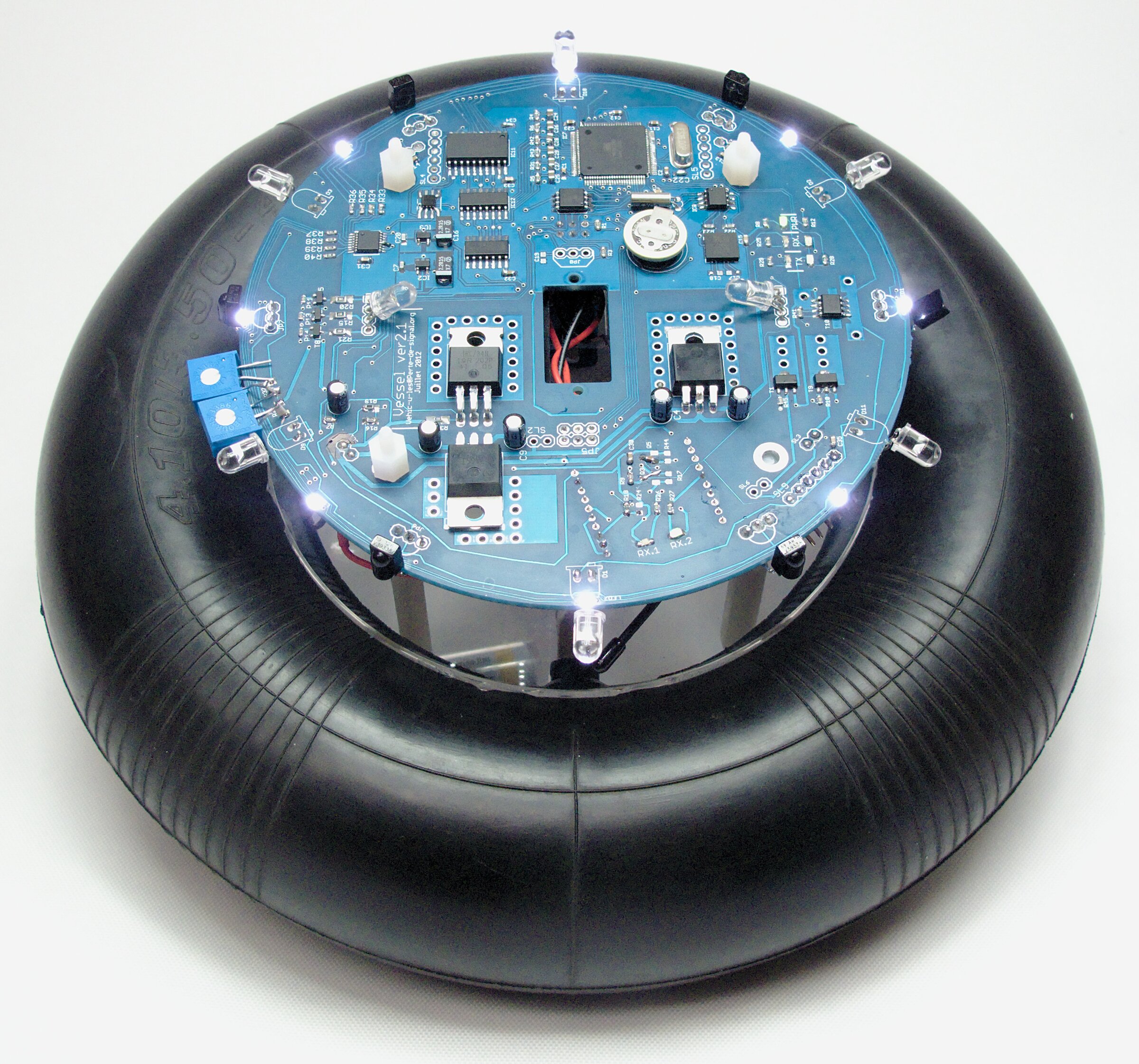}
    \caption{\textit{Vessels} custom circuit detail with microcontroller, environmental sensors, motor drivers, and RGB LEDs. Infrared diodes and receivers around the circumference allow robots to communicate with each other. Photo credit: Samuel St-Aubin.}
    \label{fig:custom_hardware}
\end{figure}

The works presented here exemplify this convergence between research-creation, engineering, and Artificial Life. Hardware and software components of these systems were all designed from scratch by the artists (e.g. Figure \ref{fig:custom_hardware}). In all three installations, aesthetic and exhibition constraints directly shaped technical innovation and \emph{vice versa}. For example, the systems involve autonomous trial-and-error learning robots designed specifically to avoid damaging or dangerous situations during long-term open-ended interaction. The circular and spherical morphologies of \textit{Vessels} and \textit{Morphosis} allow robots to explore and interact while preventing entanglement, falling, or immobilization, while in \textit{Open-Ended Ensemble}, custom magnetic slip joints permit robotic tentacles to twist into complete entanglement while limiting torque before catastrophic mechanical failure can occur. These engineering solutions emerged directly from the practical and aesthetic requirements of exhibiting adaptive robotic systems in public environments over extended durations. Simultaneously, the installations generate forms of experiential and embodied knowledge that are difficult to access through simulation or technical evaluation alone, allowing audiences to fully experience processes of emergence, adaptation, and autonomous interaction as lived and observable phenomena unfolding in real time.


\section{Audience Considerations}
The spatiotemporal, material, physical, environmental, and social conditions in which these works unfold deeply contribute to the overall aesthetic experience as much, if not more, than their technological characteristics. A foggy pond in a park, a darkened room in a museum, a crowded gallery space during a \textit{vernissage}: each introduces its own constraints and affordances, shaping how the audience perceives the work as well as how the robots behave and adapt. Here we examine how three dimensions interact to produce these encounters: 1) The presentation contexts that stage human-machine relationships; 2) The temporality through which adaptation becomes observable; and 3) The material design choices that render learning processes sensible.

\subsection{Contexts of Presentation}

These works rest on indeterminacy and surprise: they evolve, learn, and adapt in front of the audience. As such, they resist immediate intelligibility: their behaviors lie \textit{beyond human understanding}~\citep{Audry2021-Art}. To be appreciated, they must be lingered with and adopted. Like the artists who created them, the viewers must also engage in \textit{apprivoisement} with the agents, allowing their curiosity to give way to familiarity, and then, potentially, to forms of empathy, projection, and dialogue. 

In open public contexts such as \textit{Vessels}, engagement emerges from incidental encounters: passers-by are drawn in by their curiosity for the strange robotic visitors of public places. The fragility of the robots, who evolve in an uncontrolled environment and whose circuitry is precariously exposed to the elements, can subtly elicit a form of empathy, prompting conversations between people, even with strangers (see Figure \ref{fig_sim}). 
In contrast, more controlled environments such as  \textit{Morphosis} establish the conditions for sustained and focused attention. Here, the audience is asked to remove their footwear, enter a dimly lit space, and receive a brief introduction from a mediator before experiencing the 15-20 minute long robotic performance. Practically, the removal of shoes leaves the audience both vulnerable and captive, compelling them to remain within the experience. Symbolically, such gestures mark a ritualistic crossing of a threshold, emphasizing a sense of respect, akin to entering a sacred place, such as a temple or a theater.

\subsection{Timescales of Adaptation} 

\begin{figure}[!htbp]
\centering
\includegraphics[width=.95\linewidth]{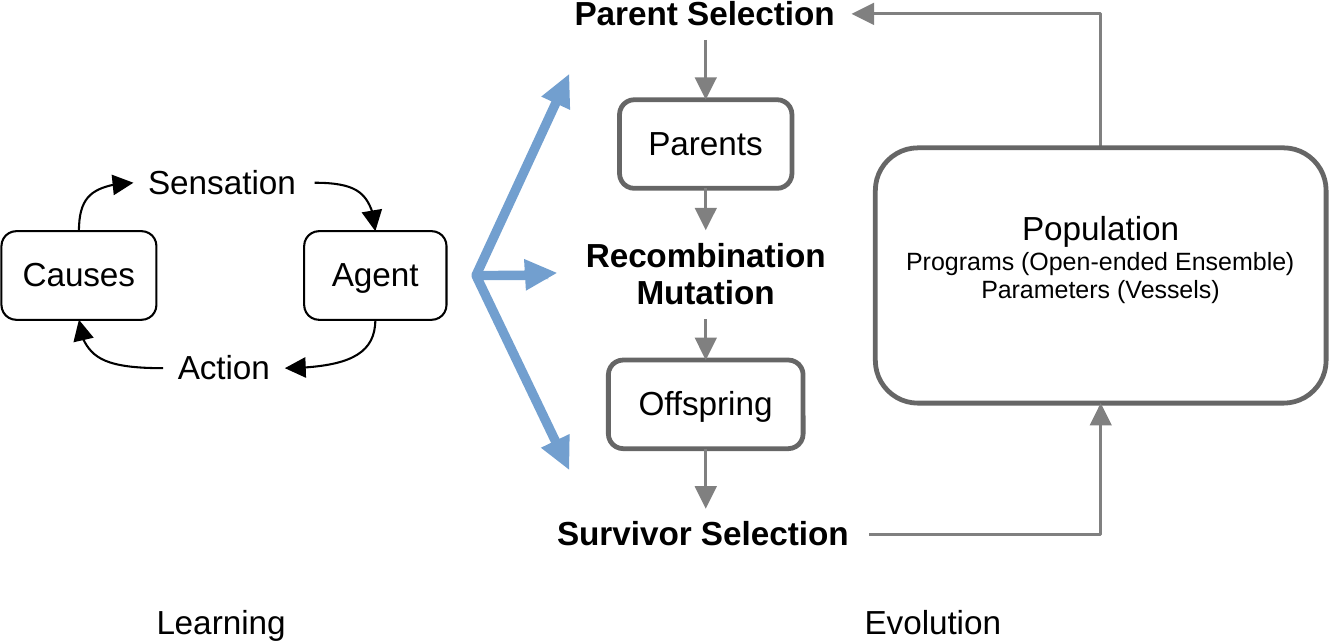}
\caption{Adaptation at two timescales in evolutionary reinforcement learning. On the Left, an individual agent interacts with its environment through a sensorimotor interface. It learns to predict the effect of its actions in the environment throughout its lifetime. This \textit{lifetime learning} loop is how adaptation occurs in \textit{Morphosis}. In \textit{vessels} and \textit{Open-ended Ensemble}, each agent's situated and subjective experience in the learning environment also influences selection and variation operators in an evolutionary adaptation cycle, shown on the Right.}
\label{fig:EvoRL}
\end{figure}

These robotic artworks are not designed to solve particular tasks, but rather to outwardly represent their underlying system. In this context, \textit{time} becomes a critical mechanism through which a system's story can be expressed. These works model natural adaptation on at least two timescales: evolution and lifetime learning (Figure \ref{fig:EvoRL}), each providing the viewer with a unique mode of experiential access to the algorithm. 

In \textit{Vessels}, each robot represents an individual within a population that evolves in real time over the course of a day or evening. Evolutionary selection and variation occur asynchronously, triggered by the robots' spatial interactions with other agents, and produce instantly perceptible behavioral changes. Thus, the work presents a real-time (non-generational) evolutionary ecosystem in which adaptations are fluid \citep{stanley:aaai06}. 

By contrast, \textit{Open-Ended Ensemble} implements a generational genetic algorithm that shifts the focus from the population to the individual. Here, an embedded computer in each robot stores an entire population of candidate programs. At any moment in time, only one program is selected from each population to be evaluated on each robot, i.e. only the \textit{Learning} component of Figure \ref{fig:EvoRL} is fully exposed to the visitor, for two individuals at a time. After all individuals in both coevolving populations have been evaluated, selection and variation occur in a single synchronous step, which updates the entire population simultaneously. This has two significant differences from \textit{Vessels}: First, it emphasizes the unique behavioral characteristics of two interacting individuals, shifting the viewer's focus from population dynamics to the adaptions in their individual \textit{lifetime} learning. Second, since only two evaluations are performed at a time, evaluating an entire population can take hours, and generational (evolutionary) adaptations are only perceptible over days or weeks. Thus, witnessing the evolutionary timescale in \textit{Open-Ended Ensemble} demands repeated visits to the gallery. Long-timescale work of this nature emphasizes local community by intrinsically rewarding visitors who share space and interact daily with the work. 

\textit{Morphosis} similarly presents adaptation through lifetime learning, but compresses the process into a timescale of minutes rather than hours, days, or weeks. While multiple robots learn simultaneously, the emphasis is less on collective behavior than on how different morphologies adapt under the same conditions. To remain perceptible within a typical gallery visit, each behavior unfolds over roughly 2--5 minutes, after which all learning parameters are reinitialized to a zero state, restarting the process from scratch. These temporal constraints required low-dimensional state and action spaces and simple reward functions, yet still produced rich and unpredictable dynamics through the entanglement of algorithmic processes with the agents' morphology, material properties, and surrounding environment.

\subsection{Algorithm-Material Entanglement}

The creation of adaptive robotic artworks requires designing a system that affords both robotic learning and human interpretation. As artificial life, these works model natural living systems as a means of better understanding or discovering something new about the natural world. A key question is how to expose the underlying systems and research questions to audience sensory experience? Material choices in system design can be exploited to provide hints to the viewer about the nature of the underlying algorithm. 

\begin{figure}[!htbp]
    \centering
    \includegraphics[width=.95\linewidth]{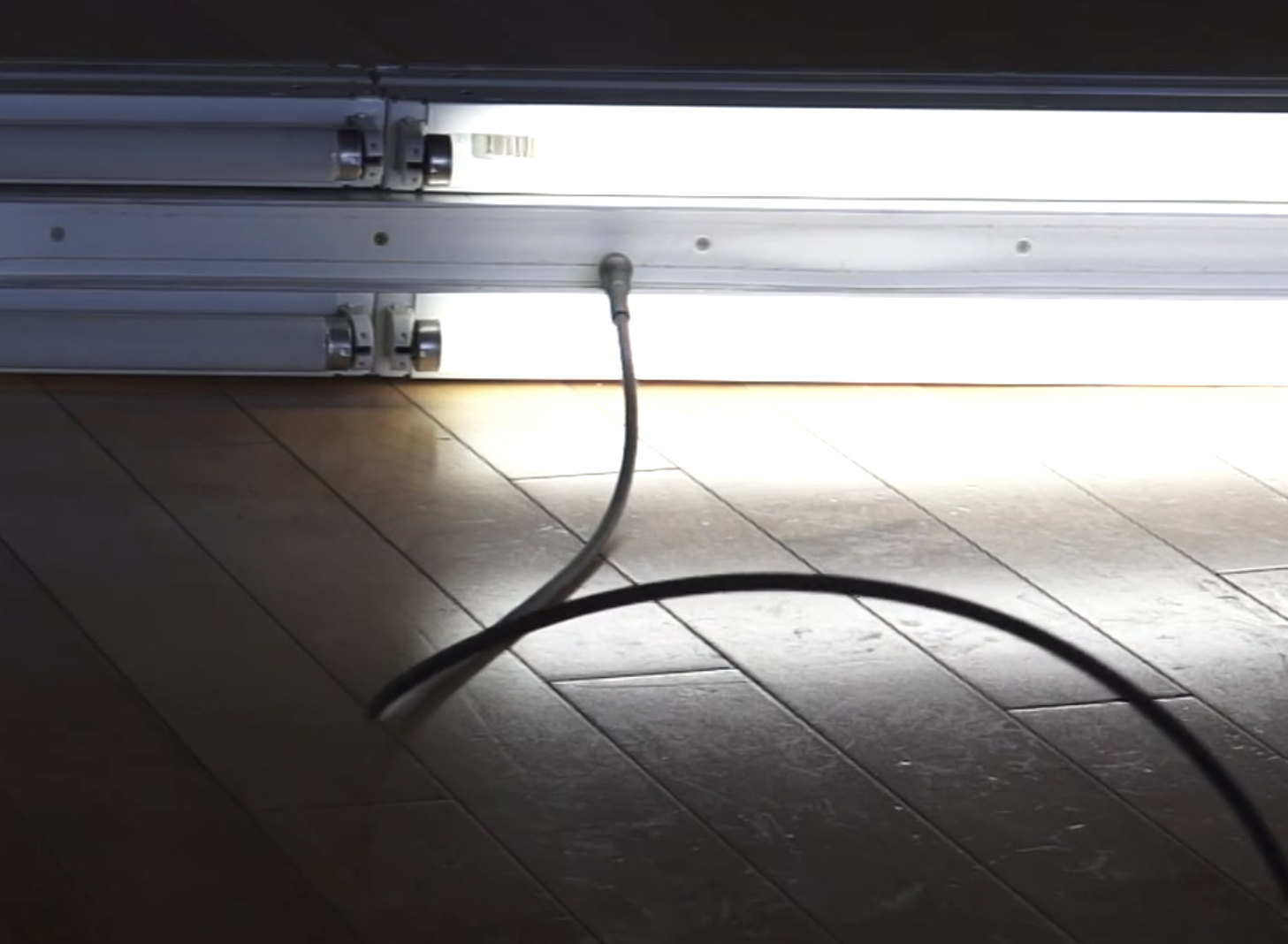}
    \caption{\textit{Open-Ended Ensemble} (detail) by Stephen Kelly (2016). View of a robotic magnetic probe crawling along a light fixture to explore a rich world of inductive hums. Photo credit: Caitlin Sutherland.}
    \label{fig:oee_close}
\end{figure}

For example, the twisting guitar cable which forms \textit{Open-Ended Ensemble}'s magnetic probe (Figure \ref{fig:oee_close}) evokes tentacles or worm-like organisms. This life-like physical property is undeniable even under test conditions, when controlled by a simple rotational force without any intelligent responsive control.

\begin{figure}[!htbp]
    \centering
    \includegraphics[width=.95\linewidth]{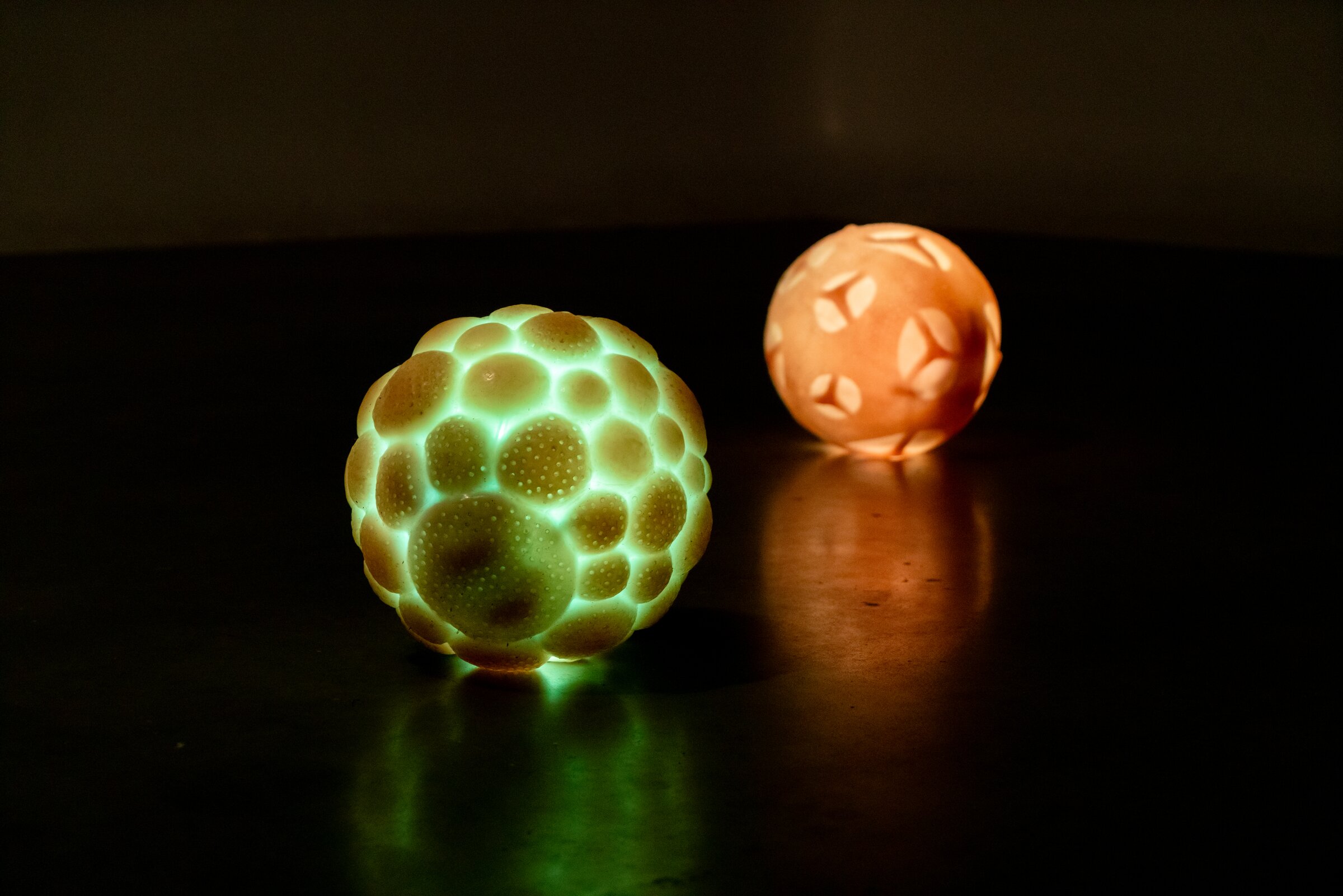}
    \caption{\textit{Morphosis} by Sofian Audry and Rosalie D. Gagné (2024). View of two robots showing details of the silicon skins and illumination. Photo credit: Léa Martin.}
    \label{fig:morphosis_close}
\end{figure}

Likewise, the morphology, color, and illumination of the spheroid entities in \textit{Morphosis} (Figure \ref{fig:morphosis_close}) are designed to convey a strong impression of life. Each of the robots is covered in a silicone skin which vary in texture and form, yielding subtle differences in both their appearance and their physical dynamics and directly contributing to their behavior, in particular by impacting their learning capabilities. Internally, an LED lighting system serves as an expressive channel: color pulses to suggest inner activity, and changes in hue mark the agent’s reinforcement signal (green for reward, red for punishment). This illumination, diffused through the semi-translucent skin and shadowed by the mechanical interior, gives viewers an intuitive cue into the agent’s ongoing adaptation. Thus, while materials define the sensorimotor interface of the robots, they also act as an evocative bridge between the human audience and the internal processes of the machine.

\begin{figure}[!htbp]
    \centering
    \includegraphics[width=.95\linewidth]{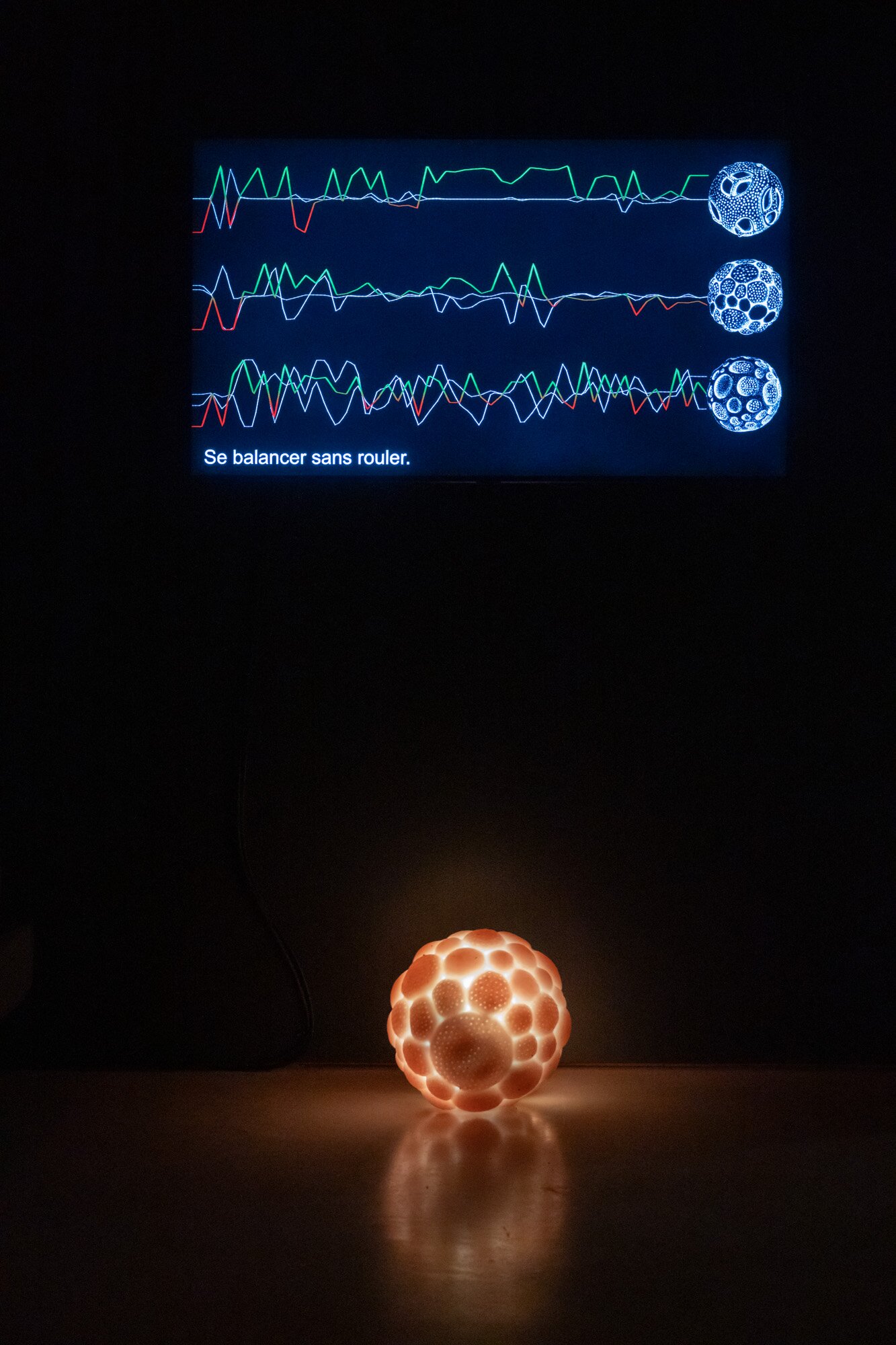}
    \caption{\textit{Morphosis} by Sofian Audry and Rosalie D. Gagné (2024). View of the visualization monitor comparing the live learning process of the robots during behavior labeled "Rocking without rolling" in which robots are positively rewarded for generating internal motion, and negatively rewarded for displacing horizontally on the floor. Photo credit: Musée de Joliette.}
    \label{fig:morphosis-visualization}
\end{figure}

In addition to the embodied cues, \textit{Morphosis} incorporates a live visualization projected in the installation space to make the learning process more intelligible, Figure \ref{fig:morphosis-visualization}. It displays each robot's adaptive trajectory: rewards are shown as a colored curve shifting from red (negative) to green (positive), directly mirroring the hue of the robot’s internal illumination. Thin white lines represent the robot’s observation state variables; these traces are unlabeled and abstract, offering no explicit legend, yet they serve as a visual proxy for what the robot perceives. 

\section{Conclusion}

Through three artistic use cases, we explore the challenges and opportunities inherent in creating robotic artworks that learn and evolve in real time, their trials and errors expressed in immersive environments that expose viewers to their underlying adaptive systems. A core dimension of working with such systems lies in the use of machine learning or evolutionary processes not as a means to optimize a specific behavior, but as a system to be experience for its own sake. We emphasize the \textit{aesthetic} value of adaptation on multiple timescales, namely evolution and learning. We show how both the creative process and the aesthetic outcomes bring into play a set of tensions between humans and machines. From our own practice of designing such generative adaptive behaviors, we deploy aesthetic strategies to bring these works to audiences, navigating between predictability and surprise, intelligibility and indeterminacy, human authorship and machine autonomy. The resulting experience is not the product of an immaterial algorithmic procedure, but of a dynamic ensemble of both physical and computational elements that includes sensors, actuators, materials, morphology, internal models, and environment. These works challenge conventional understandings of artificial intelligence as purely efficient or functional by focusing on the process rather than the objective. The artistic intention and meaning is shifted from \textit{optimization} to collective, human-machine \textit{search} and discovery.

\vfill


\printbibliography

@inproceedings{stanley:aaai06,
title={Real-Time Evolution of Neural Networks in the NERO Video Game},
author={Kenneth O. Stanley and Bobby D. Bryant and Igor Karpov and Risto Miikkulainen},
booktitle={Proceedings of the Twenty-First National Conference on Artificial Intelligence (AAAI-2006)},
address={Boston, MA},
publisher={Meno Park, CA: AAAI Press},
pages={1671--1674},
url="http://nn.cs.utexas.edu/?stanley:aaai06",
year={2006}
}

@proceedings{soros_2024,
    author = {Soros, L. B. and Adams, Alyssa M. and Kalonaris, Stefano and Witkowski, Olaf and Guckelsberger, Christian},
    title = {On Creativity and Open-Endedness},
    volume = {ALIFE 2024: Proceedings of the 2024 Artificial Life Conference},
    series = {ALIFE 2022: The 2022 Conference on Artificial Life},
    pages = {60},
    year = {2024},
    month = {07},
    doi = {10.1162/isal_a_00789},
    url = {https://doi.org/10.1162/isal\_a\_00789},
    eprint = {https://direct.mit.edu/isal/proceedings-pdf/isal2024/36/60/2461215/isal\_a\_00789.pdf},
}

@ARTICLE{pugh_2016,
AUTHOR={Pugh, Justin K.  and Soros, Lisa B.  and Stanley, Kenneth O. },     
TITLE={Quality Diversity: A New Frontier for Evolutionary Computation},
JOURNAL={Frontiers in Robotics and AI},
VOLUME={Volume 3 - 2016},
YEAR={2016},
URL={https://www.frontiersin.org/journals/robotics-and-ai/articles/10.3389/frobt.2016.00040},
DOI={10.3389/frobt.2016.00040},
ISSN={2296-9144}}

@book{herzog_2014,
 author    = {Herzog, Lena},
 title     = {Strandbeest: The Dream Machines of {Theo Jansen}},
 year      = {2014},
 publisher = {Taschen},
 address   = {Cologne}
}

@Article{Kemp_1998,
author={Kemp, Martin},
title={Latham's life-forms},
journal={Nature},
year={1998},
month={Feb},
day={01},
volume={391},
number={6670},
pages={849-849},
issn={1476-4687},
doi={10.1038/36010},
url={https://doi.org/10.1038/36010}
}

@Book{sherlock_2018,
author={Sherlock, Diana
and McKeough, Rita
and {\&} Production Society, Emmedia Gallery},
title={Rita McKeough : works},
year={2018},
publisher={EMMEDIA Gallery {\&} Production Society},
address={Calgary, Alberta},
isbn={9780986736926; 0986736929},
language={English}
}

@incollection{butts_2022,
  author    = {Shannon Butts},
  title     = {Critical Making},
  booktitle = {Keywords in Design Thinking: A Lexical Primer for Technical Communicators and Designers},
  publisher = {The WAC Clearinghouse; University Press of Colorado},
  year      = {2022},
  pages     = {1-20},
  chapter   = {12},
}

@article{kelly_2018b,
    author = {Kelly, Stephen and Heywood, Malcolm I.},
    title = {Emergent Solutions to High-Dimensional Multitask Reinforcement Learning},
    journal = {Evolutionary Computation},
    volume = {26},
    number = {3},
    pages = {347-380},
    year = {2018},
    month = {09},
    issn = {1063-6560},
    doi = {10.1162/evco_a_00232},
    url = {https://doi.org/10.1162/evco\_a\_00232},
    eprint = {https://direct.mit.edu/evco/article-pdf/26/3/347/1552375/evco\_a\_00232.pdf},
}

@inproceedings{kelly_2023,
  title={Discovering Adaptable Symbolic Algorithms from Scratch},
  author={Stephen Kelly and Daniel S. Park and Mitchell McIntire and Pranav Nashikkar and Wolfgang Banzhaf and Kalyanmoy Deb and Jie Tan and Esteban Real},
  booktitle={2023 IEEE/RSJ International Conference on Intelligent Robots and Systems (IROS)},
  year={2023},
  pages={11174-11181}
}

@book{adami_2024,
  title={The Evolution of Biological Information: How Evolution Creates Complexity, from Viruses to Brains},
  author={Adami, Christoph},
  year={2024},
  publisher={Princeton University Press}
}

@book{callebaut_2005,
    author = {Callebaut, Werner and Rasskin-Gutman, Diego and Institute, Konrad Lorenz},
    title = {Modularity: Understanding the Development and Evolution of Natural Complex Systems},
    publisher = {The MIT Press},
    year = {2005},
    month = {05},
    isbn = {9780262269698},
    doi = {10.7551/mitpress/4734.001.0001},
    url = {https://doi.org/10.7551/mitpress/4734.001.0001},
}

@Article{watson_2024,
AUTHOR = {Buckley, Christopher L. and Lewens, Tim and Levin, Michael and Millidge, Beren and Tschantz, Alexander and Watson, Richard A.},
TITLE = {Natural Induction: Spontaneous Adaptive Organisation without Natural Selection},
JOURNAL = {Entropy},
VOLUME = {26},
YEAR = {2024},
NUMBER = {9},
ARTICLE-NUMBER = {765},
URL = {https://www.mdpi.com/1099-4300/26/9/765},
PubMedID = {39330098},
ISSN = {1099-4300},
DOI = {10.3390/e26090765}
}

@inproceedings{Audry2020-Behaviour,
  ids = {Audry2020-Behavioura,Audry2020-Behaviourb},
  title = {Behaviour {{Aesthetics}} of {{Reinforcement Learning}} in a {{Robotic Art Installation}}},
  booktitle = {4th {{NeurIPS Workshop}} on {{Machine Learning}} for {{Creativity}} and {{Design}}},
  author = {Audry, Sofian and Dumont-Gagné, Rosalie and Scurto, Hugo},
  date = {2020-12},
  location = {Vancouver, Canada},
  url = {https://hal.archives-ouvertes.fr/hal-03100907},
  urldate = {2021-01-29}
}

@book{Audry2021-Art,
  title = {Art in the {{Age}} of {{Machine Learning}}},
  author = {Audry, Sofian},
  date = {2021-10},
  series = {Leonardo},
  publisher = {MIT Press},
  location = {Cambridge, MA},
  isbn = {978-0-262-04618-3},
  pagetotal = {256}
}

@incollection{Audry2024-Choreomata,
  title = {Choreomata},
  booktitle = {Choreomata: Performance and Performativity after {{AI}}},
  author = {Audry, Sofian},
  editor = {Trillo, Roberto Alonso and Poliks, Marek},
  date = {2024},
  edition = {First edition},
  pages = {283--307},
  publisher = {CRC Press, Taylor \& Francis Group},
  location = {Boca Raton},
  isbn = {978-1-003-31233-8},
  langid = {english}
}

@article{Penny1987-Simulation,
  title = {Simulation, {{Digitization}}, {{Interaction}}: {{The}} Impact of Computing on the Arts},
  author = {Penny, Simon},
  date = {1987},
  journaltitle = {Artlink},
  volume = {7},
  number = {3--4}
}

@incollection{vonUexkull1957-Stroll,
  title = {A {{Stroll Through}} the {{Worlds}} of {{Animals}} and {{Men}}},
  booktitle = {Instinctive {{Behavior}}: {{The Development}} of {{Modern Concept}}},
  author = {von Uexküll, Jakob},
  editor = {Schiller, Claire H.},
  options = {useprefix=true},
  date = {1957-06},
  pages = {5--80},
  publisher = {Intl Universities Pr Inc},
  isbn = {0-8236-2880-9}
}

@book{Whitelaw2004-Metacreation,
  title = {Metacreation: {{Art}} and {{Artificial Life}}},
  shorttitle = {Metacreation},
  author = {Whitelaw, Mitchell},
  date = {2004-03-01},
  publisher = {The MIT Press},
  location = {Cambridge, MA},
  isbn = {0-262-23234-0},
  pagetotal = {296}
}

@inproceedings{Zhao2020-SimtoReal,
  title = {Sim-to-{{Real Transfer}} in {{Deep Reinforcement Learning}} for {{Robotics}}: A {{Survey}}},
  shorttitle = {Sim-to-{{Real Transfer}} in {{Deep Reinforcement Learning}} for {{Robotics}}},
  booktitle = {2020 {{IEEE Symposium Series}} on {{Computational Intelligence}} ({{SSCI}})},
  author = {Zhao, Wenshuai and Queralta, Jorge Peña and Westerlund, Tomi},
  date = {2020-12-01},
  eprint = {2009.13303},
  eprinttype = {arXiv},
  eprintclass = {cs},
  pages = {737--744},
  doi = {10.1109/SSCI47803.2020.9308468.},
  url = {http://arxiv.org/abs/2009.13303},
  urldate = {2025-08-15}
}

@article{Brooks1986-Robust,
  title = {A Robust Layered Control System for a Mobile Robot},
  author = {Brooks, R.},
  date = {1986-03},
  journaltitle = {IEEE Journal on Robotics and Automation},
  volume = {2},
  number = {1},
  pages = {14--23},
  issn = {0882-4967},
  doi = {10.1109/JRA.1986.1087032}
}

@inproceedings{Isla2005-Handling,
  title = {Handling {{Complexity}} in the {{Halo}} 2 {{AI}}},
  booktitle = {Game {{Developer}}’s {{Conference}} 2005 {{Proceedings}}},
  author = {Isla, Damián},
  date = {2005},
  publisher = {CMP Inc.},
  location = {San Francisco, CA},
  eventtitle = {Game {{Developer}}’s {{Conference}}}
}

@inproceedings{Marzinotto2014-Unified,
  title = {Towards a Unified Behavior Trees Framework for Robot Control},
  booktitle = {2014 {{IEEE International Conference}} on {{Robotics}} and {{Automation}} ({{ICRA}})},
  author = {Marzinotto, A. and Colledanchise, M. and Smith, C. and Ogren, P.},
  date = {2014-05},
  pages = {5420--5427},
  doi = {10.1109/ICRA.2014.6907656},
  eventtitle = {2014 {{IEEE International Conference}} on {{Robotics}} and {{Automation}} ({{ICRA}})}
}

@book{Sutton1998-Reinforcement,
  title = {Reinforcement Learning: An Introduction},
  shorttitle = {Reinforcement Learning},
  author = {Sutton, Richard S. and Barto, Andrew G.},
  date = {1998},
  series = {Adaptive Computation and Machine Learning},
  publisher = {MIT Press},
  location = {Cambridge, MA},
  isbn = {978-0-262-19398-6},
  pagetotal = {322}
}

@book{Glaveanu2014-Distributed,
  title = {Distributed {{Creativity}}: {{Thinking Outside}} the {{Box}} of the {{Creative Individual}}},
  shorttitle = {Distributed {{Creativity}}},
  author = {Glăveanu, Vlad Petre},
  date = {2014},
  series = {{{SpringerBriefs}} in {{Psychology}}},
  publisher = {Springer International Publishing},
  location = {Cham},
  doi = {10.1007/978-3-319-05434-6},
  url = {https://link.springer.com/10.1007/978-3-319-05434-6},
  urldate = {2025-01-21},
  isbn = {978-3-319-05433-9 978-3-319-05434-6},
  langid = {english}
}

@article{lehman_2011,
author = {Lehman, Joel and Stanley, Kenneth O.},
title = {Abandoning objectives: Evolution through the search for novelty alone},
year = {2011},
issue_date = {Summer 2011},
publisher = {MIT Press},
address = {Cambridge, MA, USA},
volume = {19},
number = {2},
issn = {1063-6560},
url = {https://doi.org/10.1162/EVCO_a_00025},
doi = {10.1162/EVCO_a_00025},
journal = {Evol. Comput.},
month = jun,
pages = {189–223},
numpages = {35}
}

@article{pennock_2007,
author = {Robert T. Pennock},
title = {Models, simulations, instantiations, and evidence: the case of digital evolution},
journal = {Journal of Experimental \& Theoretical Artificial Intelligence},
volume = {19},
number = {1},
pages = {29--42},
year = {2007},
publisher = {Taylor \& Francis},
doi = {10.1080/09528130601116113}
}

@inproceedings{Colton2012-Computational,
  title = {Computational Creativity: The Final Frontier?},
  shorttitle = {Computational Creativity},
  booktitle = {Proceedings of the 20th {{European Conference}} on {{Artificial Intelligence}}},
  author = {Colton, Simon and Wiggins, Geraint A.},
  year = 2012,
  month = aug,
  series = {{{ECAI}}'12},
  pages = {21--26},
  publisher = {IOS Press},
  address = {NLD},
  urldate = {2026-02-03},
  isbn = {978-1-61499-097-0}
}

@inproceedings{McCormack2019-Autonomy,
  title = {Autonomy, {{Authenticity}}, {{Authorship}} and {{Intention}} in {{Computer Generated Art}}},
  booktitle = {Computational {{Intelligence}} in {{Music}}, {{Sound}}, {{Art}} and {{Design}}},
  author = {McCormack, Jon and Gifford, Toby and Hutchings, Patrick},
  editor = {Ek{\'a}rt, Anik{\'o} and Liapis, Antonios and Castro Pena, Mar{\'i}a Luz},
  year = 2019,
  pages = {35--50},
  publisher = {Springer International Publishing},
  address = {Cham},
  doi = {10.1007/978-3-030-16667-0_3},
  isbn = {978-3-030-16667-0},
  langid = {english}
}

@article{Watkins1992-Qlearning,
  title = {Q-Learning},
  author = {Watkins, Christopher J. C. H. and Dayan, Peter},
  date = {1992-05},
  journaltitle = {Machine Learning},
  shortjournal = {Mach Learn},
  volume = {8},
  number = {3--4},
  pages = {279--292},
  issn = {0885-6125, 1573-0565},
  doi = {10.1007/BF00992698},
  url = {http://link.springer.com/10.1007/BF00992698},
  urldate = {2026-02-04},
  langid = {english}
}

@article{Froese2009-Enactive,
  title = {Enactive Artificial Intelligence: {{Investigating}} the Systemic Organization of Life and Mind},
  shorttitle = {Enactive Artificial Intelligence},
  author = {Froese, Tom and Ziemke, Tom},
  date = {2009},
  journaltitle = {Artificial Intelligence},
  volume = {173},
  number = {3--4},
  pages = {466--500},
  issn = {0004-3702; 0004-3702},
  doi = {10.1016/j.artint.2008.12.001},
  url = {http://dx.doi.org/10.1016/j.artint.2008.12.001},
  urldate = {2026-02-05}
}

@thesis{Yazici2018-Relevance,
  type = {mathesis},
  title = {The {{Relevance}} of {{Umwelt Theory}} in {{Embodied Artificial Intelligence Research}}},
  author = {Yazıcı, Halil İbrahim},
  date = {2018-01-01},
  url = {https://www.academia.edu/37544202/The_Relevance_of_Umwelt_Theory_in_Embodied_Artificial_Intelligence_Research},
  urldate = {2026-02-05}
}

@article{Chapman2012-ResearchCreation,
  title = {Research-{{Creation}}: {{Intervention}}, {{Analysis}} and "{{Family Resemblances}}"},
  shorttitle = {Research-{{Creation}}},
  author = {Chapman, Owen B. and Sawchuk, Kim},
  date = {2012-04-13},
  journaltitle = {Canadian Journal of Communication},
  shortjournal = {CJC},
  volume = {37},
  number = {1},
  issn = {07005-3657},
  url = {http://www.cjc-online.ca/index.php/journal/article/view/2489}
}

@book{Loveless2019-How,
  title = {How to Make Art at the End of the World: A Manifesto for Research-Creation},
  shorttitle = {How to Make Art at the End of the World},
  author = {Loveless, Natalie},
  date = {2019},
  publisher = {Duke University Press},
  location = {Durham},
  doi = {10.1215/9781478004646},
  isbn = {978-1-4780-0402-8 978-1-4780-0464-6},
  pagetotal = {176}
}

@article{Paquin2020-Degager,
  title = {Dégager Des Connaissances de Sa Recherche-Création},
  author = {Paquin, Louis-Claude},
  date = {2020},
  journaltitle = {Louis-Claude Paquin},
  url = {https://scholar.google.com/scholar?cluster=5691163753471047436&hl=en&oi=scholarr},
  urldate = {2026-02-05}
}

@book{boden_1991,
author = {Boden, Margaret A.},
title = {The creative mind: myths and mechanisms},
year = {1991},
isbn = {0465014526},
publisher = {Basic Books, Inc.},
address = {USA}
}

@book{Brown2001-Biotica,
  title = {Biotica: Art, Emergence and Artificial Life},
  shorttitle = {Biotica},
  author = {Brown, Richard and Aleksander, Igor and MacKenzie, Jonathan and Faith, Joe},
  year = 2001,
  publisher = {RCA CRD Research},
  address = {London},
  isbn = {1-874175-33-0 978-1-874175-33-9},
  langid = {english}
}

@article{Ikegami2013-Design,
  title = {A {{Design}} for {{Living Technology}}: {{Experiments}} with the {{Mind Time Machine}}},
  shorttitle = {A {{Design}} for {{Living Technology}}},
  author = {Ikegami, Takashi},
  year = 2013,
  month = jun,
  journal = {Artificial Life},
  volume = {19},
  number = {3/4},
  pages = {387--400},
  issn = {10645462},
  doi = {10.1162/ARTL_a_00113},
  urldate = {2026-05-07},
  langid = {english}
}

@incollection{McCormack2009-Evolution,
  title = {The {{Evolution}} of {{Sonic Ecosystems}}},
  booktitle = {Artificial Life Models in Software},
  author = {McCormack, Jon},
  editor = {Komosinski, Maciej and Adamatzky, Andrew},
  year = 2009,
  edition = {2. ed},
  pages = {393--414},
  publisher = {Springer},
  address = {London},
  isbn = {978-1-84882-285-6 978-1-84882-284-9},
  langid = {english}
}

@inproceedings{Penny2009-Art,
  title = {Art and {{Artificial Life}} -- {{A Primer}}},
  booktitle = {Proceedings of the {{Digital Arts}} and {{Culture Conference}}},
  author = {Penny, Simon},
  year = 2009,
  publisher = {University of California},
  address = {Irvine}
}

@article{Penny2010-Twenty,
  title = {Twenty Years of Artificial Life Art},
  author = {Penny, Simon},
  year = 2010,
  month = sep,
  journal = {Digital Creativity},
  volume = {21},
  number = {3},
  pages = {197--204},
  issn = {1462-6268, 1744-3806},
  doi = {10.1080/14626261003654640},
  urldate = {2017-06-28},
  langid = {english}
}

@inproceedings{Schmidhuber1991-Curiosity,
  title = {A Possibility for Implementing Curiosity and Boredom in Model-Building Neural Controllers},
  booktitle = {Proceedings of the International Conference on Simulation of Adaptive Behavior: {{From}} Animals to Animats},
  author = {Schmidhuber, Jürgen},
  date = {1991},
  pages = {222--227},
  publisher = {MIT Press}
}

@article{Sommerer1999-Art,
  title = {Art as a {{Living System}}: {{Interactive Computer Artworks}}},
  shorttitle = {Art as a {{Living System}}},
  author = {Sommerer, Christa and Mignonneau, Laurent},
  year = 1999,
  journal = {Leonardo},
  volume = {32},
  number = {3},
  pages = {165--173},
  issn = {0024-094X},
  urldate = {2018-05-07}
}

@article{Tenhaaf1998-Art,
  title = {As {{Art Is Lifelike}}: {{Evolution}}, {{Art}}, and the {{Readymade}}},
  shorttitle = {As {{Art Is Lifelike}}},
  author = {Tenhaaf, Nell},
  year = 1998,
  journal = {Leonardo},
  volume = {31},
  number = {5},
  pages = {397--404},
  issn = {0024-094X},
  doi = {10.2307/1576605},
  urldate = {2018-05-07}
}

@article{Tenhaaf2008-Art,
  title = {Art {{Embodies A-Life}}: {{The VIDA Competition}}},
  shorttitle = {Art {{Embodies A-Life}}},
  author = {Tenhaaf, Nell and Gaetano, Paula and Cadet, France and Muelas, Federico and Draves, Scott and Teran, Michelle and Mann, Jeff and Nishijima, Haruki and Verstappen, Mar{\'i}a and Driessens, Erwin and B{\"o}hlen, Marc and Rinker, J. T.},
  year = 2008,
  journal = {Leonardo},
  volume = {41},
  number = {1},
  pages = {6--24},
  issn = {0024-094X},
  urldate = {2018-05-07}
}

@article{Wu2024-Survey,
  title = {A {{Survey}} of {{Recent Practice}} of {{Artificial Life}} in {{Visual Art}}},
  author = {Wu, Zi-Wei and Qu, Huamin and Zhang, Kang},
  year = 2024,
  month = feb,
  journal = {Artificial Life},
  volume = {30},
  number = {1},
  pages = {106--135},
  issn = {1064-5462},
  doi = {10.1162/artl_a_00433},
  urldate = {2026-05-07}
}

\end{document}